\documentclass[12pt]{article}

\usepackage[margin=1.25in]{geometry}
\usepackage{graphicx}
\usepackage{amsthm}
\usepackage{amsfonts}
\usepackage{amssymb}
\usepackage{bm}
\usepackage{amsmath}
\usepackage{color}
\usepackage{microtype}
\usepackage{enumitem}
\usepackage[toc,page]{appendix}
\usepackage{relsize}
\usepackage{lineno}
\usepackage{multirow}
\usepackage{changepage}
\newtheorem{theorem}{Theorem}
\newtheorem{assumption}{Assumption}

\newtheorem{lemma}{Lemma}

\newtheorem{definition}{Definition}

\theoremstyle{definition}
\newtheorem{remark}{Remark}
\graphicspath{ {FigsEtc/} }

\begin{document}

\thispagestyle{empty}

\begin{center}
{\LARGE The Existence of Equilibrium Flows}

\bigskip
\bigskip

Alfred Galichon$^{\dag }$, Larry Samuelson{\small $^{\flat }$}, and Lucas Vernet{\small $^{\S }$}

\bigskip

September 1, 2022

\bigskip

\end{center}

\noindent {\bf Abstract~~}  
Galichon, Samuelson and Vernet (2022) introduced a class of problems, {\em equilibrium flow problems}, that nests several classical economic models such as bipartite matching models, minimum-cost flow problems and hedonic pricing models. We establish conditions for the existence of equilibrium prices in the equilibrium flow problem, in the process generalizing Hall's \cite{Hall1935} theorem.

\bigskip
\bigskip

\bigskip

{\small
	\tableofcontents}

\vfill
$^{\dag }$Economics Department, FAS, and Mathematics Department, Courant Institute, New York University; and Economics Department, Sciences Po. Email: ag133@nyu.edu.  Galichon gratefully acknowledges funding from ERC grant CoG-866274.

{\small $^{\flat }$}Department of Economics, Yale University. Email: larry.samuelson@yale.edu.

{\small $^{\S }$}Economics Department, Sciences Po, and Banque de France. Email: 

\noindent lucas.vernet@acpr.banque-france.fr.

\clearpage


\setcounter{page}{1}

\begin{center}
{\LARGE The Existence of Equilibrium Flows}
\end{center}

\section{Introduction}

Galichon, Samuelson and Vernet \cite{GalSamVer2022} introduced a class of problems, {\em equilibrium flow problems}, that nests several classical economic models such as bipartite matching models, minimum-cost flow problems and hedonic pricing models.  This paper establishes conditions for the existence of equilibrium prices in the equilibrium flow problem, generalizing Hall's \cite{Hall1935} theorem.

\section{Equilibrium Flow Problem}

This section duplicates the presentation of the equilibrium flow problem in Galichon, Samuelson and Vernet \cite{GalSamVer2022}.

\subsection{Formulation}

\paragraph{Network.} Consider a network $\left(\mathcal{Z},\mathcal{A}\right)$ where $%
\mathcal{Z}$ is a finite set of nodes and $\mathcal{A} \subseteq \mathcal Z \times \mathcal Z$ is the set of directed
arcs.  If $xy \in \mathcal A$, we say that $xy$ is the arc from $x\in \mathcal Z$ to $y\in \mathcal Z$, and we say that $x$ is the starting point of the arc, while $y$ is its end point. We assume that there is no arc in $\mathcal A$ whose starting point coincides with the end point. We describe the network with an $|\mathcal{A}|\times
|\mathcal{Z}|$ \emph{arc-node incidence matrix} matrix $\nabla$, defined by letting,
for $xy\in \mathcal{A}$ and $z\in Z$
\[
\nabla _{xy,z}=\mathbf 1_{\left\{ z=y\right\}} -\mathbf 1_{\left\{ z=x\right\}}.
\]
We thus have $\nabla _{xy,z}=1$ if $xy$ is an arc  ending at $z$, and $\nabla _{xy,z}=-1$
if $xy$ is an arc beginning at $z$.  Otherwise $\nabla _{xy,z}=0$.

\paragraph{Prices, connection functions.} Let $p\in\mathbb{R}^{\mathcal{Z}}$ be a price vector, where we interpret $p_{z}$ as the price at node $z$.  
To have a concrete description, though we do not require this interpretation, one may consider a trader operating on node $xy$, who is able to purchase one unit of a commodity at node $x$, ship it along arc $xy$ toward node $y$, and resell it at node $y$. Given the resale price at node $y$, there is a certain threshold value of the price at node $x$ such that the trader is indifferent between engaging in the trade or not. This value is an increasing and continuous function of the price at node $y$, and can be expressed as $G_{xy}(p_y)$, where for each arc $xy \in \mathcal A$, the function 
$G_{xy}: \mathbb R \to \mathbb R$ is continuous and increasing, and called the \emph{connection function}.\footnote{This is related to the idea of a Galois connections, which explains the choice of the letter $G$;  see~N\"{o}ldeke and Samuelson \cite{NandS2017}.}
Hence, if $p_x>G_{xy}(p_y)$, the purchase price at node $x$ is excessive, and the trader will not engage in the trade. On the contrary, if $p_x<G_{xy}(p_y)$, the purchase price is strictly below indifference level and positive rent can be made from the trade on the arc $xy$.

Our framework allows for any situation where the per-unit rent $\Pi_{xy}(p_x,p_y)$ of the trade on arc $xy$ is a continuous and possibly nonlinear function of $p_x$ and $p_y$, increasing in the resale price $p_y$ and decreasing in the purchase price $p_x$. In that case, $G_{xy}(p_y)$ is implicitly defined from $\Pi_{xy}$ by $G_{xy}(p_y) = \Pi_{xy}(.,p_y)^{-1}(0)$, or equivalently, $\Pi_{xy}(G_{xy}(p_y),p_y)=0$.

We refer to an equilibrium flow problem as a triple  $(\mathcal Z,\mathcal A,G)$.

\paragraph{Exiting flow, internal flow, mass balance.} Let $q\in \mathbb R^{\mathcal Z}$ with $\sum _{z \in \mathcal Z}q_z=0$ attach a net flow to each node $z\in \mathcal Z$.  If $q_z>0$, then the net quantity $|q_z|$ must flow into node $z$, while $q_z<0$ indicates that the net quantity $|q_z|$ must flow away from node $z$. Hence, we call $q$  the vector of \emph{exiting flows}. We let  $\mu\in\mathbb R^{\mathcal A}_+$ be the vector of \emph{internal flows} along arcs, so that  $\mu_{xy}$ is the flow through arc $xy$.  The feasibility condition connecting these notions is that, for any $z\in\mathcal{Z}_{0}$, the total internal flow that arrives at $z$ minus the total internal flow that leaves $z$ equals the exiting flow at $z$, that is 
\begin{equation*}
	\sum_{x:xz\in\mathcal{A}}\mu_{xz}-\sum_{y:zy\in\mathcal{A}}\mu_{zy}=q_{z},
\end{equation*}
which we call the \emph{mass balance equation}, and which can be rewritten as
\begin{equation}\label{guitar}
	\nabla^{\intercal}\mu=q.
\end{equation}
The interpretation of these flows will depend on the application of the equilibrium flow problem.  The flows may represent quantities of commodities, volumes of traffic, assignments of objects, matches of individuals, and so on.

\subsection{Equilibrium}

\paragraph{Equilibrium flow.} The triple $(q,\mu,p)\in \mathbb R^{\mathcal Z}\times \mathbb R^{\mathcal A}_+\times \mathbb R^{\mathcal Z}$ is an equilibrium outcome if it satisfies three conditions.  The first condition is the conservation of the flow given by the mass balance equation~\eqref{guitar}.  The second condition is that there  there is no positive rent on any arc, that is:
\begin{equation*}
	p_x \geq G_{xy}\left(p_y\right)\ ~\forall xy\in\mathcal{A}.
\end{equation*}
Our third condition is that arcs with negative rents carry no flow.   Hence $\mu_{xy}>0\implies  p_x \leq G_{xy}\left(p_y\right)$,
which combines with no-positive-rent requirement to yield
$\mu_{xy}>0\implies p_x = G_{xy}\left(p\right)$. This is a complementary
slackness condition, which can be written
\begin{equation*}
	\sum_{xy\in\mathcal{A}}\mu_{xy} \left( p_x - G_{xy}\left(p_y\right) \right) =0.
\end{equation*}
The interpretation of these rent conditions will again depend on the application.  In some cases, they will be the counterparts of zero-profit conditions in markets with entry, while in other cases they will play the role of incentive constraints.

In summary, we define:

\begin{definition}[{\bf Equilibrium Flow Outcome}] 
	The triple $(q,\mu,p)\in \mathbb R^{\mathcal Z}\times \mathbb R^{\mathcal A}_+\times \mathbb R^{\mathcal Z}$ is an equilibrium  outcome when the following conditions are met:
	
	\bigskip
	
	(i) $\nabla^{\intercal}\mu=q$
	
	(ii) $p_x \geq G_{xy}\left(p_y\right)\ ~\forall xy\in\mathcal{A}$
	
	(iii) $\sum_{xy\in\mathcal{A}}\mu_{xy} \left( p_x - G_{xy}\left(p_y\right) \right) =0$.
\end{definition}

\noindent The first condition implies $\sum_{z\in\mathcal Z}q_z=0$.  Notice that if $p$ satisfies condition (ii), then setting $q=0$ and $\mu = 0$ ensures that the remaining conditions are satisfied. Indeed, if $(q,\mu,p)$ is a equilibrium flow outcome, then so is $(\lambda q,\lambda \mu,p)$ for any nonnegative scalar $\lambda$.  Hence, there will either be no equilibrium flow outcome (if there is no $p$ satisfying condition (ii)) or there will be multiple equilibrium flows outcomes.  

\section{Existence of Equilibrium}\label{extract}

We fix a vector of exit flows $q$ (with $\sum_{z\in\mathcal Z}q_z=0$) and ask whether there exists a price vector $p$ and flow $\mu$ such that $(q,\mu,p)$ is an equilibrium of the equilibrium flow problem.

\subsection{Feasibility Conditions}

For a subset $B$ of $\mathcal{Z}$, an arc $xy$ is said to be outward if $x\in B$ and $y\notin B$.  A  subset $B$ of $\mathcal{Z}$ is called retaining if there is no arc outward of $B$. Hence, subset $B$ is retaining  if and only if
\[
\nabla \mathbf 1_{B}\geq 0.
\]

Our existence result obviously requires some feasibility condition.  In the extreme, one cannot find an equilibrium flow in a network with no arcs (and nontrivial $q$).  We state the feasibility condition in terms of retaining sets.

\begin{assumption}\label{whale}
If $B$ is a retaining set, then $q(B) = \sum_{z\in
\mathcal{B}}q_{z} \ge 0$.
\end{assumption}

Assumption \ref{whale} serves its intended purpose of ensuring it is at least possible to construct a flow that transports the goods in the network from their origins to their destinations:%

\begin{lemma}\label{ship}
Let $q\in \mathbb{R}^{\mathcal{Z}}$ be such that $\sum_{z\in
\mathcal{Z}}q_{z}=0$. Then the following statements are equivalent.

[\ref{ship}.1] There exists $ \mu \in \mathbb{R}_{+}^{\mathcal{A}}$ such that $q=\nabla ^{\top }\mu $.

[\ref{ship}.2] For each $B$ retaining subset of $\mathcal{Z}$, one has $q\left( B\right) \geq 0$.
\end{lemma}

\paragraph{Proof}
That [\ref{ship}.1]  implies [\ref{ship}.2] is straightforward: If there is no arc from a node in $B$ to a node in $\mathcal{Z}\backslash B$, then mass can only enter into $B$, and so we must have $q\left( B\right) \geq 0$.

To show the converse, we exploit Hoffman's \cite{Hoffman1960} circulation theorem.    For any subset $B\subset \mathcal Z$, let $\mathcal{O}\left( B\right) =\left\{ xy\in \mathcal{A}:x\in B,y\notin B\right\} $, and $\mathcal{I}\left( B\right) =\left\{ xy\in \mathcal{A}:x\notin B,y\in B\right\} $ be respectively the set of outward and inward arcs of $B$.  Hoffman's theorem states that, given $q\in \mathbb{R}^{\mathcal{Z}}$ with $\sum_{z\in \mathcal Z}q_z = 0$  and $\underline{\mu }$ and $\overline{\mu }$ in $\mathbb{R}_{+}^{\mathcal{A}}$ there exists $\mu \in \mathbb{R}_{+}^{\mathcal{A}}$ such that
\[
\nabla ^{\intercal }\mu  =q ~~~~~~{\rm and}~~~~~~
\underline{\mu } \leq \mu \leq \overline{\mu }
\]
if and only if for all subsets $B$ of $\mathcal{Z}$, one has%
\begin{equation*}
\sum_{z\in B}q_{z}\leq \sum_{a\in \mathcal{I}\left( B\right) }\overline{\mu }%
_{a}-\sum_{a\in \mathcal{O}\left( B\right) }\underline{\mu }_{a}.
\end{equation*}
This result is  usually stated for $q=0$, but the adaptation to nonzero $q$ straightforward.  With this result in hand, let $\overline \mu_a = +\infty$ and $\underline \mu_a = 0$, so that we can rewrite Hoffman's condition as the requirement that for all sets $\tilde B$ that have no inbound links,
\[
q(\tilde B) \le 0.
\]
We then argue that this condition is equivalent to the second condition in Lemma \ref{ship}.  Call a set with no inbound links {\em repelling}.  The result then follows from noting that the set $B\subset \mathcal Z$ is retaining if and only if $B^c$ is repelling, and for any set $B$,
\[
q(B)+q(B^c) = 0.
\]
\hfill\rule{.125in}{.125in}

\begin{remark}[{\bf Hall's marriage theorem}]
Lemma \ref{ship} implies Hall's \cite{Hall1935} marriage theorem.   To see this, assume we are in the bipartite case to which Hall's lemma applies.  Hence,  $\mathcal{Z}=\mathcal{X}\cup \mathcal{Y}$ and $\mathcal{A}\subseteq \mathcal{X}\times \mathcal{Y}$. 

From Lemma \ref{ship}, we then have that $q$ is attainable if and only if for all retaining sets $B$, one has $q\left( B\right)
\geq 0$.  That is, for all retaining sets $B$, one has
\[
q\left( B\cap \mathcal{X}%
^{c}\right) +q\left( B\cap \mathcal{X}\right) \ge 0.
\]
Therefore for all subsets $B$ of $\mathcal{Z}\backslash \mathcal{X}$ and for all  $B^{\prime }$ in $\mathcal{X}$, then
\begin{equation}\label{whack}
[B\subset\mathcal{Z}\backslash \mathcal{X}, B'\subset \mathcal X, B\cup B' {\rm ~retaining~ }] \Longrightarrow [q\left( B\right) \geq -q\left( B^{\prime }\right)].
\end{equation}
Now let $B'\subset \mathcal X$ and let $B$ be the set of nodes $z$ in $\mathcal Y$ with the property that there is an arc from a node in $B$ to $z$.  Then by construction $B\cup B'$ is retaining, and so \eqref{whack} gives
\[
q\left( B\right) \geq -q\left( B^{\prime }\right),
\]
which is Hall's marriage theorem.\hfill\rule{.125in}{.125in}
\end{remark}

We make an additional, technical feasibility assumption. 

\begin{assumption}\label{minor}
For each $z\in \mathcal Z$, either $q_z>0$ or  there exists a path from $z$ to a
node $y\in \mathcal Z$ with $q_y>0$.
\end{assumption}

\noindent  To see why we refer to this as a technical assumption, notice that it is automatically satisfied if $q_z>0$, and is implied by Assumption \ref{whale} if $q_z<0$.  However, Assumption \ref{whale} is compatible with the existence of a node $z$ with $q_z=0$ that fails Assumption \ref{minor}.  In this case, it is impossible for an equilibrium to exhibit flow through node $z$, and we can without loss of generality eliminate node $z$ from the network.  Assumption \ref{minor} thus ensures that there are no obviously irrelevant nodes in the network.  We could accommodate such nodes, but then would require another step in our equilibrium construction, to attach prices to such nodes.

\subsection{A Profitability Condition}\label{puffin}

Assumption \ref{whale} ensures that there exists a flow capable of moving goods from their sources to their destinations.  However, in equilibrium the flow must make use of only zero-rent arcs and there must be no positive-rent arcs. We accordingly require a condition on the profitability of arcs.

For $x$ and $y$ in $\mathcal{Z}$, we introduce the reduced connection $%
\tilde{G}_{xy}\left( p_{y}\right) $ as%
\begin{equation*}
\tilde{G}_{xy}\left( p_{y}\right) =\sup_{\substack{ %
x_{0}=x,x_{1},x_{2},...,x_{k+1}=y  \\ x_{\ell}x_{\ell+1}\in \mathcal{A}}}%
G_{xx_{1}}\circ G_{x_{1}x_{2}}\circ ...\circ G_{x_{k}y}\left( p_{y}\right)
\end{equation*}%
which can either be finite or $+\infty $ if this quantity is unbounded. We
shall take the convention that $\tilde{G}_{xy}\left( p_{y}\right) =-\infty $
if there is no directed path between $x$ and $y$.

 We say that the collection of nodes $(x_0,x_1,\ldots x_{k+1})$ is a loop if  it contains $k+1$ distinct nodes and $x_0=x_{k+1}=: x$.  
We assume:

\begin{assumption}\label{major}
If $(x_0,x_{1},\ldots,x_{k+1})$ is a
loop, then $p>G_{xx_{1}}\circ G_{x_{1}x_{2}}\circ ...\circ G_{x_{k}x}\left(
p\right) $ for all $p$.
\end{assumption}

Assumption \ref{major} ensures there is no way to achieve positive rents by moving flow through a loop, with no net change in the allocation.  If such a loop existed, continual movement along this loop would constitute a money pump, giving rise to infinite rents.  Equivalently, if Assumption \ref{major} fails, then it may be impossible to assign prices to nodes in such a way as to ensure the absence of positive-rent arcs.

\subsection{The Existence Result}

Lemma \ref{ship} ensures that it is possible to transport the required mass from sources to targets.   Assumption \ref{major} ensures that it is possible to attach prices to nodes so that there are no positive-rent arcs.  In equilibrium, we require some coordination between these two constructions, in that equilibrium flow must transport the required mass from sources to targets using only zero-rent arcs.  It would accordingly be no surprise if addition conditions were required to ensure the existence of an equilibrium flow.  However, these assumptions suffice:

\begin{theorem}\label{chant}
Let Assumptions \ref{whale}-\ref{major} hold.  Then there exists an equilibrium flow.
\end{theorem}

The proof of this result, occupying Sections \eqref{bob}--\eqref{dylan}, associates the equilibrium flow problem with an associated bipartite matching problem, and then exploits existence results for the bipartite matching problem.  This association may be useful for other purposes.

\subsection{Equilibrium Flows and Bipartite Solutions}\label{bob}

Given an equilibrium flow problem, define the source nodes as $\mathcal{X}=\left\{ x\in \mathcal{Z}:q_{x}<0\right\} $ and the target nodes as $\mathcal{Y}=\left\{ y\in \mathcal{Z}:q_{y}>0\right\} $.  We now associate with the equilibrium flow problem $(\mathcal Z,\mathcal A, G)$ the bipartite imperfectly transferable utility matching problem $(\mathcal X, \mathcal Y, \tilde G)$.

\begin{lemma}\label{candy}
Fix a profile of exit flows $q$ with $\sum_{z\in\mathcal Z}q_z=0$.  Then there exits an equilibrium $(q,\mu,p)$ for the equilibrium flow problem $(\mathcal Z,\mathcal A, G)$ if and only if there exists an equilibrium $(q,\tilde \mu,\tilde p)$ for the associated bipartite problem $(\mathcal X, \mathcal Y, \tilde G)$ such that $p$ and $\tilde p$ agree on $\mathcal X$ and $\mathcal Y$.
\end{lemma}

\paragraph{Proof.}
First, let $\left(q, \mu ,p\right) $ be a solution to the equilibrium flow problem. We derive a corresponding bipartite solution.

First, we have for $x\in \mathcal{X}$ and $y\in \mathcal{Y}$ that
\begin{equation*}
p_{x}\geq \tilde{G}_{xy}\left( p_{y}\right) .
\end{equation*}
Indeed, for any path $x=x_{0},x_{1},x_{2},...,x_{k+1}=y$ from $x$ to $y$, we
have (from the no-positive-rent equilibrium condition) $p_{x_{k}}\geq G_{x_{k}x_{k+1}}\left( p_{x_{k+1}}\right) $ and thus
\begin{equation*}
p_{x}\geq G_{xx_{1}}\circ G_{x_{1}x_{2}}\circ ...\circ G_{x_{k}y}\left(
p_{y}\right).
\end{equation*}
Taking the maximum over paths from $x$ to $y$, we have $p_{x}\geq \tilde{G}_{xy}\left(p_{y}\right) $.

By the flow decomposition theorem, one can write
\begin{equation*}
\mu =\sum_{p\in \Pi }\mu ^{p}+\sum_{l\in L}\mu ^{l}
\end{equation*}%
where $\mu ^{p}$ is the flow associated with path $p$, and $\mu ^{l}$ is the flow associated with loop $l$, $\Pi $ is the set of paths, and $L$ is the set of loops.  Clearly $\mu_{l}=0$ for any loop $l$, otherwise the assumption of absence of profitable loop would be violated.

Consider $\Pi _{xy}$, which is the set of paths beginning at $x$ and ending at $y$.
Let $\tilde\mu _{xy}=\sum_{p\in \Pi _{xy}}\mu^{p}$. It is straightforward that for all $x\in \mathcal X$ and $y\in \mathcal Y$ we have
\begin{equation*}
\begin{array}{c}
\sum_{x\in \mathcal{Y}}\tilde\mu _{xy}=n_{x} \\
\sum_{x\in \mathcal{X}}\tilde\mu _{xy}=m_{y}.
\end{array}
\end{equation*}
%
%

%
%
%
%

In addition, we have that  $\tilde \mu _{xy}>0$ implies that there is a path  $(x_{0},x_{1},x_{2},...,x_{k+1})$ from $x$ ($=x_0$) to $y$ ($=x_{k+1}$) along which the flow is
strictly positive. Thus, along this path, $p_{x_{\ell}}=G_{x_{\ell}x_{\ell+1}}\left(
p_{x_{\ell+1}}\right) $ and thus
\begin{equation*}
p_{x}=G_{xx_{1}}\circ G_{x_{1}x_{2}}\circ ...\circ G_{x_{k}y}\left(
p_{y}\right);
\end{equation*}%
but as $p_{x}\geq \tilde{G}_{xy}\left( p_{y}\right) \geq G_{xx_{1}}\circ G_{x_{1}x_{2}}\circ ...\circ G_{x_{k}y}\left( p_{y}\right) $, it follow that equality holds everywhere and thus
\begin{equation*}
p_{x}=\tilde{G}_{xy}\left( p_{y}\right) .
\end{equation*}
Therefore we have the an equilibrium $(q,\tilde\mu,\tilde p)$ for the bipartite problem, i.e., satisfying:
\begin{equation}\label{shellfish}
\begin{array}{l}
\sum_{x\in \mathcal{Y}}\tilde\mu _{xy}=n_{x} \\
\sum_{x\in \mathcal{X}}\tilde\mu _{xy}=m_{y} \\
\tilde p_{x}\geq \tilde{G}_{xy}\left(\tilde p_{y}\right) \\
\tilde\mu _{xy}>0\implies \tilde p_{x}=\tilde{G}_{xy}\left( \tilde p_{y}\right)%
\end{array}.
\end{equation}

For the reverse direction of the proof, suppose $(q,\tilde\mu,\tilde p)$ is an equilibrium of the bipartite solution $\left( p_{x}^{\ast },p_{y}^{\ast }\right) $, and hence satisfies \eqref{shellfish}.  To find an associated equilibrium $(q,\mu,p)$ of the equilibrium flow problem, we retain the specification of $q$, let $\mu$ be any flow consistent with $\tilde \mu$, and  define an iterative procedure that generates the prices $p$.  For the initial step, we let
\[
p_{z}^{0}=
\left\{
\begin{array}{ll}
\tilde p_{z} &{\rm if}~z\in \mathcal{X}\cup
\mathcal{Y}\\
-\infty &{\rm otherwise}
\end{array}
\right.,
\]
with the induction step given by
\[
p_{x}^{t+1}=\max \left\{ p_{x}^{t},\max_{y:xy\in \mathcal{A}}\left\{
G_{xy}\left( p_{y}^{t}\right) \right\} \right\} .
\]

We then complete the argument with the the following lemma:

\begin{lemma}
for $t$ large enough (but finite), $(q,\mu,p^{t})$ an equilibrium of the
equilibrium flow problem.
\end{lemma}

\paragraph{Proof.}
We establish this result in a series of steps:

\begin{enumerate}

\item  After a finite number of steps, $p_{z}^{t}>-\infty
$ for all $z$.

To see this, let $d\left( x\rightsquigarrow y\right)$ be the length of the shortest path from $x$ to $y$.  Then we have  $p_{x}^{t}=\max_{y\in \mathcal{Y}%
}\max_{x\rightsquigarrow y:l\left( x\rightsquigarrow y\right) \leq
t}G_{x\rightsquigarrow y}\left( p_{y}\right) $.  Hence, if there is a directed
path of length at most $k$ from $z$ to $\mathcal{Y}$, then the price of $z$
will become finite after in at most $k$ steps.
As a result, all prices will be updated after%
\begin{equation*}
\max_{z}\min_{y\in \mathcal{Y}}d\left( z,y\right),
\end{equation*}%
and so  after a finite number of step, $p_{z}^{t}>-\infty
$ for all $z$.

\item  The price of a node $y\in \mathcal{Y}$ is never updated by the
algorithm.

Assume that one updates $y\in \mathcal{Y}$ at some step. Then there is a
chain $y\rightsquigarrow y^{\prime }$ from $y$ to some $y^{\prime }\in
\mathcal{Y}$ with $p_{y}<G_{y\rightsquigarrow y^{\prime }}\left(
p_{y^{\prime }}\right) $. Let $x\in \mathcal{X}$ be an element of $\mathcal{X%
}$ such that $\pi _{xy}>0$. One has $p_{x}=\tilde{G}_{xy}\left( p_{y}\right)
$, thus $p_{x}<\tilde{G}_{xy}\circ G_{y\rightsquigarrow y^{\prime }}\left(
p_{y^{\prime }}\right) =G_{x\rightsquigarrow y^{\prime }}\left( p_{y^{\prime
}}\right) \leq \tilde{G}_{xy^{\prime }}\left( p_{y^{\prime }}\right) $, a
contradiction.

\item  There exists a T such that for all $t>T$ the algorithm is
stationary .

We construct an updating digraph as follows: at $t=0$, the graph has no
edge.  At $t>0$, if $p_{x}^{t+1}<\max_{y:xy\in \mathcal{A}}\left\{ G_{xy}\left(
p_{y}^{t}\right) \right\} $, then we pick up one $y$ that attains the maximum
(resolving ties arbitrarily) and we add a directed arc from $x$ to $y$. It
is a direct consequence of Assumption \ref{major} that this digraph is a forest.
Because there are only a finite number of possible forests, the updating
process must stop within finite many steps.

\item   For all $x\in X$, $p_{x}^{T}=p_{x}^{\ast }$.

By contradiction, if there exists a $x$ such that $p_{x}^{T}\neq p_{x}^{\ast
}$ when the algorithm is stationary then:

\begin{itemize}
\item If $p_{x}^{T}>p_{x}^{\ast }$, then there exists a path $%
x\rightsquigarrow y$ such that $p_{x}^{T}=G_{x\rightsquigarrow y}\left(
p_{y}\right) >p_{x}^{\ast }=\max_{y^{\prime }\in \mathcal{Y}}\tilde{G}%
_{xy^{\prime }}\left( p_{y^{\prime }}\right) >G_{x\rightsquigarrow y}\left(
p_{y}\right) $, \ contradiction.

\item If $p_{x}^{T}<p_{x}^{\ast }$, then there is a $y\in \mathcal{Y}$ and a
path $x\rightsquigarrow y$ such that $p_{x}^{\ast }=G_{x\rightsquigarrow
y}\left( p_{y}\right) $. Moving along path $x\rightsquigarrow y$, take the
last $z$ such that $p_{z}^{T}\neq \tilde{G}_{zy}\left( p_{y}\right) $, and $%
z^{\prime }$ be its successor. One has by definition $p_{z}^{T}<\tilde{G}%
_{zy}\left( p_{y}\right) =G_{zz^{\prime }}\left( \tilde{G}_{z^{\prime
}y}\left( p_{y}\right) \right) =G_{zz^{\prime }}\left( p_{z^{\prime
}}^{T}\right) $, which contradicts the fact that the algorithm has reached
stationarity.
\end{itemize}
\end{enumerate}
\hfill\rule{.125in}{.125in}

\noindent This completes the proof of Lemma \ref{candy}.\hfill\rule{.125in}{.125in}

\subsection{Bipartite Existence Result}

\begin{lemma}\label{sharp}
If the bipartite problem satisfies Hall's condition, then there exists an equilibrium flow for the bipartite network.
\end{lemma}

\paragraph{Proof.}
Let $n_{x}$ and $m_{y}$ be two vectors on $\mathcal{X}$ and $\mathcal{Y}$
which sum to the same amount.  Our interpretation is that $n_x$ identifies the mass at each node $x\in X$ that must be transported away from $x$, and $n_y$ identifies the mass at each node $y\in Y$ that must be transported to $y$.  The requirement that these vectors have the same sum is equivalent to the condition $\sum_{z\in Z}q_z=0$ in the equilibrium flow problem.

Consider $\mathcal{A}\subseteq \mathcal{X}\times \mathcal{Y}$, and assume
Hall's conditions are satisfied: for all $K\subseteq \mathcal{X}$,
\[
\sum_{x\in K}n_{x}\geq \sum_{y\in \mathcal{Y}}m_{y}\mathbf 1_{\left\{ \forall x:xy\in
\mathcal{A},x\in K\right\}} .
\]
This condition would be automatically satisfied if the network is fully connected, i.e., if $\mathcal A = \mathcal X \times \mathcal Y$, but we allow the network to be incomplete.

The strategy is to exploit existence results for complete bipartite networks.  Toward this end, we fix an integer $n>0$, we extend the set of arcs $\mathcal A$ to the set $\mathcal X\times \mathcal Y$ by letting, for each $x\in (\mathcal X\times \mathcal Y)\setminus \mathcal A$
\[
G_{xy}(p_y) = p_y-n.
\]
We now consider a matching on the complete graph.  If this matching directs flow along the arc $x\in (\mathcal X\times \mathcal Y)\setminus \mathcal A$, then we must have $p_x=p_y-n$.  We then argue that for sufficiently large $n$, flow along any such arc is so unprofitable that the equilibrium will not exhibit such flow.  This will ensure that the equilibrium is also an equilibrium of the problem with the set of arcs $\mathcal A$.

Hence, let $\left( \mu _{xy}^{E},p_{x},p_{y}\right) $ be such that

\bigskip

(i) $\sum_{y}\mu _{xy}^{E}=n_{x}$ and $\sum_{x}\mu _{xy}^{E}=m_{y}$,

(ii) $p_{x}\ge G_{xy}\left( p_{y}\right) $ for all $x\in \mathcal{X}$, $y\in
\mathcal{Y}$

(iii) $\mu _{xy}^{E}>0$ implies $p_{x}=G_{xy}\left( p_{y}\right) $ if $xy\in
\mathcal{A}$, and $p_{x}=p_{y}-n$ otherwise.

(iv) $p_{0}=0$ for some ground node $0\in \mathcal{X\cup Y}$

\bigskip

\noindent The existence of this matching follows from Proposition 6 of N\"{o}ldeke and Samuelson.  The key to applying this proposition is to note that the function $G_{xy}(p_y)$ is increasing, continuous, and as $p_y$ varies, sweeps out the entire real line.

Consider a directed graph on $\mathcal{Z}=\mathcal{X}\cup \mathcal{Y}$ whose
set of arcs is
\[
\mathcal{A}^{\ast }=\left\{ xy\in \mathcal{A}\right\} \cup \left\{ yx:x\in
\mathcal{X},y\in \mathcal{Y},\mu _{xy}^{E}>0\right\}
\]

We show the following intermediate result:

\begin{lemma}
 Assume $x^{\ast }y^{\ast }\notin \mathcal{A}$ and $\mu _{x^{\ast
}y^{\ast }}^{E}>0$. There exists no partition $\mathcal{Z}$ into $A$ and $B$
in such a way that $y^{\ast }\in A$, $x^{\ast }\in B$ and there are no arcs
of $\mathcal{A}^{\ast }$ from $B$ to $A$.
\end{lemma}

\paragraph{Proof.}
We show the lemma by contradiction. Suppose such partition existed, and
denote $A^{\mathcal{X}}=A\cap \mathcal{X}$, $A^{\mathcal{Y}}=A\cap \mathcal{Y%
}$, and $B^{\mathcal{X}}$ and $B^{\mathcal{Y}}$ in a similar fashion.

Because there are no arcs from $B^{\mathcal{X}}$ to $A^{\mathcal{Y}}$, Hall's
conditions imply that $n_x\left( A^{\mathcal{X}}\right) \geq m_y\left( A^{%
\mathcal{Y}}\right) $.

Now, because there are no arcs from $B^{\mathcal{Y}}$ to $A^{\mathcal{X}}$,
the existence of $\mu ^{E}$ implies that $n_x\left( B^{\mathcal{X}}\right)
\geq m_y\left( B^{\mathcal{Y}}\right) $.

Summing the previous two inequalities yields $n_x\left( \mathcal{X}\right)
\geq m_y\left( \mathcal{Y}\right) $, but as this holds as an equality, we get
an equality in the two previous inequalities.
Thus $n_x\left( A^{\mathcal{X}}\right) =m_y\left( A^{\mathcal{Y}}\right) $ and $%
n_x\left( B^{\mathcal{X}}\right) =m_y\left( B^{\mathcal{Y}}\right) $.

Now $n_x\left( A^{\mathcal{X}}\right) =\sum_{x\in A^{\mathcal{X}}}\sum_{y\in
\mathcal{Y}}\mu _{xy}^{E}$, but for $x\in A^{\mathcal{X}}$ and $y\in B^{%
\mathcal{Y}}$, $\mu _{xy}^{E}=0$ otherwise there would be an arc of $%
\mathcal{A}^{\ast }$ from $B$ to $A$. Hence%
\[
n_x\left( A^{\mathcal{X}}\right) =\sum_{x\in A^{\mathcal{X}}}\sum_{y\in A^{%
\mathcal{Y}}}\mu _{xy}^{E}
\]
Comparing with
\[
n_Y\left( A^{\mathcal{Y}}\right) =\sum_{x\in \mathcal{X}}\sum_{y\in A^{%
\mathcal{Y}}}\mu _{xy}^{E}
\]%
yields a difference of%
\[
\sum_{x\in B^{\mathcal{X}}}\sum_{y\in A^{\mathcal{Y}}}\mu _{xy}^{E},
\]%
and hence for all $x\in B^{\mathcal{X}}$ and $y\in A^{\mathcal{Y}}$, $\mu
_{xy}^{E}=0$, which contradicts $x^{\ast }\in B^{\mathcal{X}}$, $y^*\in
\mathcal{A}^{Y}$ and $\mu _{x^*y^*}^{E}>0$. This contradiction proves the claim.
\hfill\rule{.125in}{.125in}
\bigskip

The proof of Lemma \ref{sharp} is then completed by the following lemma.

\begin{lemma}\label{phone}
For $n$ large enough, $\mu _{x^{\ast }y^{\ast }}^{E}>0$ implies $%
x^{\ast }y^{\ast }\in \mathcal{A}$.
\end{lemma}

\paragraph{Proof.}
We prove the result in the following manner. Start by taking $A=\left\{ y^{\ast
}\right\} $ and $B=\mathcal{Z}\backslash \left\{ y^{\ast }\right\} $. Then
there must be an arc from $B$ to $A$. Call $z$ the origin point of that arc,
and repeat with $A=\left\{ y^{\ast },z\right\} $ and $B=\mathcal{Z}%
\backslash \left\{ y^{\ast },z\right\} $, and so on, until $x^{\ast }$ is chosen.
We end up with a path from $x^{\ast }$ to $y^{\ast }$. Calling $%
z_{0}=y^{\ast },z_{1},...,z_{K}=x^{\ast }$ that path, we have that $%
p_{y^{\ast }}\le G_{y^{\ast }z_{1}}\circ ...\circ G_{z_{K-1}z_{K}}\left(
p_{x^{\ast }}\right) $ where for $yx\in \mathcal{A}^{\ast }\backslash
\mathcal{A}$, one has defined $G_{yx}=G_{xy}^{-1}$. However, $\mu _{x^{\ast
}y^{\ast }}^{E}>0$ would imply $p_{x^{\ast }}=G_{x^{\ast }y^{\ast }}\left(
p_{y^{\ast }}\right) =p_{y^{\ast }}-n$.  Hence, we have
\begin{eqnarray}
p_{y^{\ast }}&\le&G_{y^{\ast }z_{1}}\circ ...\circ G_{z_{K-1}z_{K}}\left(
p_{x^{\ast }}\right)\label{red}\\
p_{y^*} &=& p_{x^*}+n.\label{sox}
\end{eqnarray}
Equation (\ref{sox}) shows that as $n$ grows, $p_{y^*}-p_{x^*}$ diverges.
We show that this gives a contradiction, establishing the theorem.
Write $p_{y^*}-p_{x^*}$ as
\begin{eqnarray*}
p_{y^*}-p_{x^*}&=& p_{y^*}-p_{z_1}~~~~~~~~~(1)\\
&+& p_{z_1}-p_{z_2}~~~~~~~~~(2)\\
&+& p_{z_2}-p_{z_3}~~~~~~~~~(3)\\
&+& p_{z_3}-p_{z_4}~~~~~~~~~(4)\\
&+& p_{z_4}-p_{z_5}~~~~~~~~~(5)\\
&+& p_{z_5}-p_{z_6}~~~~~~~~~(6)\\
&\vdots&\\
&+& p_{z_K}-p_{x^*}~~~~~~~~~(\cdot).
\end{eqnarray*}
The terms with odd numbers correspond to arcs from $X$ to $Y$, and are contained in $\mathcal A$.  (The final $\cdot$ is an odd-numbered term.)  The terms with even numbers are arcs from $Y$ to $X$, contained in $\mathcal A^*\setminus \mathcal A$.  Suppose first that as $n$ grows, $p_{y^*}$ approaches a limit.  This imposes a lower bound on $p_{z_1}$ (since $z_1y^*\in \mathcal A$ and hence we need nonpositive rent on this arc).  Since flow moves along arc $z_1z_2$, this in turn ensures either that there is a lower bound on $p_{z_2}$ (if the arc $z_1z_2\in \mathcal A$) or that $p_{z_2}$ diverges to $+\infty$ (if $z_1z_2\not \in \mathcal A$).  Continuing in this fashion, it cannot be the case that $p_{x^*}$ diverges to $-\infty$, a contradiction.  A similar argument can be constructed, beginning from the assumption that $p_{x^*}$ approaches a limit as $n$ grows large, and yet a similar construction ensures that it cannot be the case that $p_{y^*}\rightarrow +\infty$ and $p_{x^*}\rightarrow -\infty$.

It remains to preclude the possibilities that $p_{x^*}$ and $p_{y^*}$ both approach $+\infty$ or both approach $-\infty$, which is done in Appendix \ref{cable}.  
this ensures we have an equilibrium involving only arcs in $\mathcal A$, which is to say that we have an equilibrium flow.
\hfill\rule{.125in}{.125in}

\bigskip

\noindent This completes the proof of Lemma \ref{sharp}.\hfill\rule{.125in}{.125in}

\subsection{Closing the Ring}\label{dylan}

The proof of Theorem \ref{chant} is then completed by the following:

\begin{lemma}
The network $(\mathcal Z,\mathcal A)$ and exit flow satisfy
\[
q(B)\ge0
\]
for all retaining $B$ if and only if the associated bipartite matching problem $(\mathcal X,\mathcal Y)$ together with the exit flow $q$ satisfies Hall's marriage condition.
\end{lemma}

\paragraph{Proof.}
First, suppose $q(B)\ge0$ for all retaining sets.  Suppose $\tilde X\cup \tilde Y$ is a subset of $X\times Y$ for which, in the bipartite matching problem, there are no arcs from nodes in $\tilde X$ to nodes outside $\tilde Y$.  In the network flow problem, define a sequence of sets by letting $Z_1 = \tilde X\cup \tilde Y$, letting $Z_2$ be the union of $Z_1$ and any node that is the endpoint of an arc whose origin lies in $Z_1$; letting $Z_2$ be the union of $Z_1$ and any node that is the endpoint of an arc whose origin lies in $Z_1$, and so on.  This sequence terminates (since the set $\mathcal Z$ of nodes is finite) in a set $Z^*$ that is retaining (otherwise the process would not terminate) and hence satisfies $q(Z)\ge 0$.  Moreover, $Z^*$ contains no node $z$ for which $q(z)>0$ that is not contained in $\tilde Y$ (since otherwise there would be a link in the bipartite matching problem from some node in $\tilde X$ to a node outside $\tilde Y$, a contradiction).  Then we have
\[
0\le q(Z^*)\le q(\tilde X\cup \tilde Y),
\]
giving Hall's condition.

Second, for the other direction, let $\tilde Z$ be a retaining set.  Then
\[
q(\tilde Z) = q((X\cap \tilde Z)\cup(Y\cap \tilde Z))
\]
Because $Z$ is retaining, in the bipartite problem there are no links from $X\cap \tilde Z$ to $Y\cap \tilde Z$.  Hence, Hall's condition gives $q((X\cap \tilde Z)\cup(Y\cap \tilde Z))$, establishing the needed result.
\hfill\rule{.125in}{.125in}

\appendix

\section{Appendix: Completion of Proof of Lemma \ref{phone}}\label{cable}

We present the argument for the case in which both approach $+\infty$, with the other case being analogous.  If $p_{x^*}$ and $p_{y^*}$ both approach $+\infty$, then we can divide  $\mathcal X$  into disjoint sets $\underline {\mathcal X}$ and $\overline {\mathcal X}$ and
can divide  $\mathcal Y$  into disjoint sets $\underline {\mathcal Y}$ and $\overline {\mathcal Y}$ such that the prices in $\overline {\mathcal X}$ and $\overline {\mathcal Y}$ diverge and
the prices in $\underline {\mathcal X}$ and $\underline {\mathcal Y}$ do not diverge.  The sets $\overline {\mathcal X}$ and $\overline {\mathcal Y}$
are nonempty, because they contain $x^*$ and $y^*$.  The set $\underline {\mathcal X}$ is nonempty because it contains the ground node $p_0$.  There must be no arcs in $\mathcal A$ from $\underline {\mathcal X}$ to $\overline {\mathcal Y}$ (since otherwise such an arc would generate positive rent, a contradiction.  There must exist no flow from nodes in $\overline {\mathcal X}$ to nodes in $\underline {\mathcal Y}$ (since these arcs earn negative rent).  Any flow from $\underline {\mathcal X}$ to $\underline {\mathcal Y}$
must occur along arcs in $\mathcal A$, since otherwise the arc must earn negative rent.  Hall's condition must then hold for the system whose nodes are $\underline {\mathcal X}$ and $\underline {\mathcal Y}$, and then must separately hold the system whose nodes are $\overline {\mathcal X}$ and $\overline {\mathcal Y}$.

Intuitively, we now repeat the previous existence argument separately for each of these systems, with one ground $\underline p_0$ for  the system whose nodes are $\underline {\mathcal X}$ and $\underline {\mathcal Y}$ and one ground node $\overline p_0$ for the system whose nodes are $\overline {\mathcal X}$ and $\overline {\mathcal Y}$.  This gives us an equilibrium for each system separately.  We cannot be sure that we have an equilibrium of the combines systems, since there may be arcs in $\mathcal A$ from $\overline {\mathcal X}$ to $\underline {\mathcal Y}$.  However, if we set $\underline p_0$ sufficiently low and $\overline p_0$ sufficiently high, these arcs will carry no flow, and we will indeed have an equilibrium.

To make this argument precise we note that given any sets $\mathcal X$ and $\mathcal Y$ satisfying Hall's condition, we can partition these sets into $((\mathcal X_1,\mathcal Y_1),(\mathcal X_2,\mathcal Y_2),\ldots,(\mathcal X_n,\mathcal Y_n))$ with the property that Hall's condition holds separately for each pair $(\mathcal X_k,\mathcal Y_k)$, and the are no arcs in $\mathcal A$ from any $\mathcal X_k$ to $\mathcal Y_j$ for $j>k$.  Now we start at the bottom and use our basic argument to fix a price at a ground node $p_0(1)$ in $\mathcal X_1$ and find an equilibrium of the system $(\mathcal X_1,\mathcal Y_1)$.  In particular, we cannot have prices diverge when we get to the last step of this argument, since otherwise we could have decomposed $(\mathcal X_1,\mathcal Y_1)$ further.  Next, we fix a ground node $p_0(2)$ in $\mathcal X_2$ and repeat the same construction in $(\mathcal X_2,\mathcal Y_2)$.  This gives us an equilibrium for $(\mathcal X_2,\mathcal Y_2)$, and moreover, by making $p_0(2)$ sufficiently large, we can ensure that there is no positive rent on any arc in $\mathcal A$ from a node in $\mathcal X_2$ to a node in $\mathcal Y_1$ (recall that by construction there are no arcs from $\mathcal X_1$ to $\mathcal Y_2$), ensuring that together, we have an equilibrium for the system consisting
of $\mathcal X_1\cup \mathcal X_2$ and $\mathcal Y_1\cup\mathcal Y_2$.%
\footnote{We are using here the property that as we increase the price $p_0(2)$, all the equilibrium prices in the system $(\mathcal X_2,\mathcal Y_2)$ increase.  If this was not the case, we could further decompose $(\mathcal X_2,\mathcal Y_2)$, which by assumption we cannot.}
We then continue in this fashion.

\bibliographystyle{plain}
\bibliography{larrybib}

\end{document}